\documentclass[prb,twocolumn,showpacs]{revtex4}
\usepackage{graphicx}

\begin{document}

\def \reofour {$\mathrm{(TMTSF)_{2}ReO_4}$\,}
\def \clofour {$\mathrm{(TMTSF)_{2}ClO_4}$\,}
\def \areofour {$\mathrm{ReO_4^-}$\,}
\def \aclofour {$\mathrm{ClO_4^-}$\,}
\def \pfsix {$\mathrm{(TMTSF)_{2}PF_6}$\,}
\def \asfsix {$\mathrm{(TMTSF)_{2}AsF_{6}}$}
\def \tmtsfx {$\mathrm{(TMTSF)_{2}X}$\,}
\def \tao {$\mathrm{T_{AO,2}(P)}$\,}
\def \taot {$\mathrm{T_{AO,3}(P)}$\,}

\title{Upper critical field divergence induced by mesoscopic phase separation in the organic superconductor  $\mathrm{(TMTSF)_{2}ReO_4}$\,}
\author{C.V.Colin}
 \altaffiliation {Present address : Department of Chemical Physics, Materials Science Center, Nijenborgh 4, 9747 AG Groningen, The Netherlands, e-mail : C.V.Colin@rug.nl}
\author{B.Salameh}
 \altaffiliation {Department of Applied Physics, Tafila Technical University, P.O.Box179, Tafila 66110, Jordan}
\author{C.R.Pasquier}%
\affiliation{Laboratoire de Physique des Solides, Univ. Paris-Sud, CNRS, UMR 8502, F-91405 Orsay, France}

\author{K.Bechgaard} 
\affiliation{Department of Chemistry, H.C.Oersted Institute, Universitetsparken 5, DK2100, Copenhagen, Denmark}

\begin{abstract} 
Due to the competition of two anion orders, $\mathrm{(TMTSF)_{2}ReO_4}$\, presents a phase coexistence between semiconducting and metallic (superconducting) regions (filaments or droplets) in a wide range of pressure. In this regime, the superconducting upper critical field for H parallel to both \textit{c*} and \textit{b'} axes present a linear part at low fields followed by a divergence above a cross-over field. This cross-over corresponds to the 3D-2D decoupling transition expected in filamentary or granular superconductors. The sharpness of the transition also demonstrates that all filaments are of similar sizes and self organize in a very ordered way. The distance between the filaments and their cross-section are estimated. 
\end{abstract}


\maketitle

Superconductivity in anisotropic and inhomogeneous materials remains an important open question especially the influence of the microstructure on the superconducting properties. In organic superconductors, there are essentially two types of inhomogeneities. On the one hand, a small substitution of \aclofour anions by \areofour in \clofour leads to a fast decrease of the critical temperature \cite{Joo04} and a phase repulsion with the competing spin density wave (SDW) state \cite{Joo05}. On the other hand, large SDW domains may be introduced by a high cooling rate in pristine \clofour \cite{Schwenk84,Pouget90} or by adjusting the pressure in \pfsix \cite{Lee02,Vuletic02}. The critical temperature is nearly independent of the SDW domain density while the upper critical field \cite{Lee02} or the critical current \cite {Vuletic02} are strongly affected. The formation of alternating slabs from the two different ground states has therefore been suggested in \pfsix to explain the data. Recently, in \reofour, we have reported a self-organization of charge in a wide range of pressure \cite{Colin06} due to the competition between the two anion orders $q_2=(1/2,1/2,1/2)$ and $q_3=(0,1/2,1/2)$\cite{Moret82}. It has been demonstrated that metallic (superconducting) droplets or filaments associated with $q_3$ order, elongated along the \textit{a}-axis are embedded in the semiconducting matrix where $q_2$ order prevails\cite{Colin06}. The existence of a possible structural ordering of the filaments can be checked from the superconducting state when the magnetic field is applied perpendicular to the filaments. Indeed, the influence of a regular arrangement of filaments on the upper critical field has been already estimated a long time ago by Turkevich \textit{et al.}\cite{Turkevich79}. A 3D-2D dimensional cross-over occurs at a characteristic temperature T* which depends only on the distance, d$_{\bot}$, between the filaments in the plane perpendicular to the magnetic field. Above T*, the normal core of the vortices penetrates the filaments whereas below T*, they stack between the filaments. This model is in fact an extension of the model applied to determine the upper critical field evolution in lamellar superconductors \cite{Klemm75} when the magnetic field is perfectly aligned parallel to the superconducting layers and was shown to adjust the data in Nb/Cu multilayers for instance \cite{Krasnov96}. Such a dimensional cross-over exists also for superconducting droplets embedded in an insulating matrix and the finite size of the droplets leads to a parabolic evolution of the upper critical field at low temperatures\cite{Deutscher80}. 

In this communication, we report, for the first time in an organic compound, a precise estimate of the microstructure of the sample in its phase separation regime. Indeed, in \reofour, we will show that due to the self-organization of charges and the quality of the samples, the dimensional cross-over appears clearly on the upper critical field line which establishes a low dispersion in the distances between the superconducting filaments. Then, from the evolution of the upper critical field at lower temperatures, we will extract an estimate of the cross-section of the filaments in the (\textit{b-c}) plane.

The experiments were performed in the pressure range 10-11.8kbar where both q2 and q3 orders coexist. The metallic parts, associated to the $q_3$ order, present a superconducting transition (T$_c$ =1.52K) which is nearly pressure independent \cite{Parkin81}. The results are presented on two different samples: on the first one,  the current and the magnetic field are applied along the \textit{c*}-axis, so that $\rho_c$ is measured. On the second one, we performed $\rho_a$ resistivity measurements with the magnetic field, H, applied nearly along the \textit{b'}-axis. The alignment was made by eye and was performed before the application of the hydrostatic pressure. The dimensions of the samples are typically $2.5\times0.2\times0.05$ mm$^{3}$.

\begin{figure}[htb]
	\centering
		\includegraphics[width=0.45\textwidth]{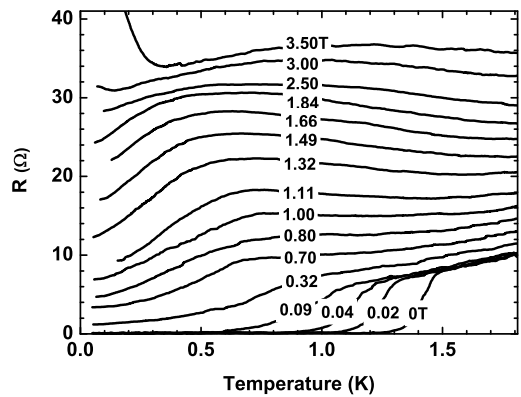}
	\caption{Resistance along the \textit{c*}-axis versus temperature in \reofour at P=10kbar and H//\textit{c*}. The curves are performed with an applied magnetic field of 0 to 3.50T from bottom to top.}
	\label{rhoc}
\end{figure}

\begin{figure}[htb]
	\centering
		\includegraphics[width=0.45\textwidth]{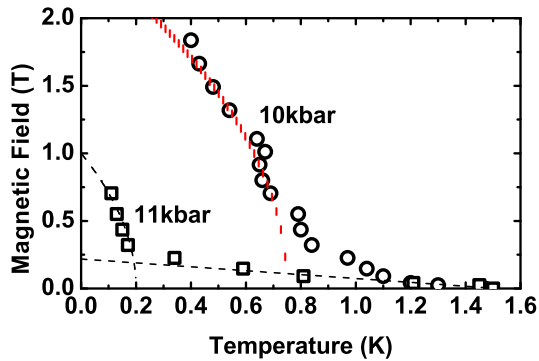}
	\caption{Temperature-magnetic field phase diagram of \reofour for P=10 kbar (open circles) and 11 kbar (open squares) and H//\textit{c*}. The straight line represents the upper critical field in the homogeneous state while the dotted lines represent fits of the $H_{c2}$ lines with a parabolic law.}
	\label{Hc2}
\end{figure}

\begin{figure}[htb]
	\centering
		\includegraphics[width=0.45\textwidth]{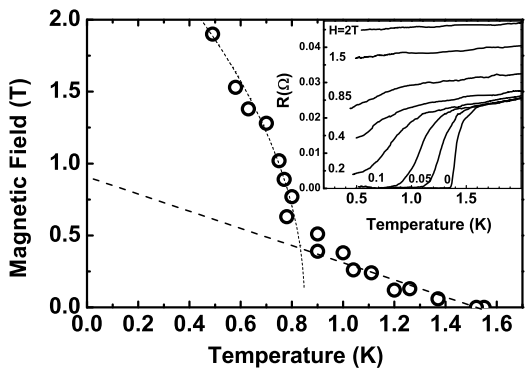}
	\caption{Temperature-magnetic field phase diagram of \reofour for P=10 kbar and H nearly parallel to \textit{b'}. The straight line represents the extrapolation of the linear part of the upper critical field while the dotted lines represents the fit of the $H_{c2}$ line with a parabolic law. Inset : resistance versus temperature for H nearly parallel to \textit{b'}.}
	\label{Hc2b}
\end{figure}

Figure \ref{rhoc} presents typical interlayer resistivity versus temperature curves obtained at 10kbar when the magnetic field is applied along the \textit{c*}-axis.  We first notice that above T$_c$ and for nearly all magnetic fields, the resistivity presents a metallic character contrary to the SDW-metal phase separation in \pfsix. Above 2T, the transition is so broad that it is nearly impossible to extract a critical temperature. Then, above 3T, the field induced spin density wave state manifests at ultra low temperature by a rapid increase of the resistance \cite{Greene86}. At each magnetic field, the critical temperature is defined by the onset temperature in a similar way to previous determination in \pfsix\cite{Lee02}. This criterion is justified since this definition also corresponds to the temperature where non linearities in the voltage-current characteristics disappear. The magnetic field-temperature phase diagram deduced from these data is shown in Fig.\ref{Hc2}. The same data can be extracted for various pressures and the phase diagram obtained for P=11kbar is also shown on this figure. At high temperatures, for all pressures, the upper critical field varies smoothly and quasi-linearly with the magnetic field. However, at low temperatures, a drastic upward curvature in the H$_{c2}$ curve is clearly visible which is translated to lower temperatures as pressure is increased. Finally, at P=12kbar, H$_{c2}$ is quasi-linear with magnetic field, no more divergence from this linear behavior is observed in the explored temperature range. Figure \ref{Hc2b} presents the magnetic field-temperature phase diagram for H//\textit{b'}, at P=10kbar, deduced from the $\rho_a$ resistivity curves obtained at different magnetic fields which are shown in the inset. Here again, just above T$_c$, the resistivity remains metallic in the whole explored range of magnetic fields. Similarly to H//\textit{c*} data, a linear variation of H$_{c2}$ is observed at low fields which is replaced by a strong upturn of the upper critical field below a characteristic temperature. 

These evolutions of the upper critical field with temperature for both field directions are in strong contrast with the reported observations in \pfsix where the upturn of the upper critical field is observed starting from the lowest fields \cite{Lee02}. The data may then be analyzed using a different model and it appears clearly that the image of a filamentary superconductor already suggested using resistivity anisotropy data is valid\cite{Colin06}. Since \reofour is strongly one dimensional, we will denote $\xi_a$, $\xi_b$ and $\xi_c$, the coherence lengths along the axes \textit{a}, \textit{b'} and \textit{c*} axes respectively. In a regular stack of droplets, filaments or slabs, a dimensional cross-over in the superconducting state is expected at a fixed temperature T* defined by :
\begin{equation}
\xi_{\bot}(\mathrm{T^*})=\frac{\xi_{\bot}(\mathrm{0})}{\sqrt{1-\frac{\mathrm{T^*}}{T_c}}}=\frac{d_{\bot}}{\sqrt{2}}
\label{eq1}
\end{equation}
where $\xi_{\bot}(\mathrm{T})$ is the temperature dependent coherence length perpendicular to the field, H, and $d_{\bot}$ is the distance between the superconducting objects in the plane perpendicular to H. Above this temperature T*, the superconductor is 3D and the upper critical field varies linearly with temperature like a standard homogeneous superconductor. Below T*, vortices remain between the filaments and are of Josephson nature \cite{Feinberg90} and a Lawrence-Doniach model may be applied\cite{LD}. In this two-dimensional regime, a divergence of the upper critical field is expected and is limited by finite-size effects, the paramagnetic limit or eventually spin-orbit coupling. The latter possibility is not pertinent in organic conductors made of light elements. However, in strongly one-dimensional systems, a Fulde-Ferrell-Larkin-Ovchinikov (FFLO) superconducting state \cite{Fulde64,Larkin64} may also be considered \cite{Dupuis95}. However, FFLO superconductivity leads to an inhomogeneous superconducting state in the reciprocal space whereas we are dealing here with phase separation in the real space. Moreover, FFLO superconductivity is expected to appear at $T_{FFLO} \approx 0.56 T_c$ in contradiction with our data at 11kbar. Finally, in this 2D regime, a square-root variation of the upper critical field line is expected below T* including the finite size of the superconducting objects\cite{Klemm75,Deutscher80}. Assuming spherical metallic grains, Deutscher \textit{et al.} estimated the upper critical field at zero temperature to be given by :

\begin{equation}
H_{c2}(0)=\sqrt{\frac{5}{3}}\frac{\Phi_0}{2\pi}\frac{1}{\xi_{\bot}(0)R}\sqrt{1-2\frac{\xi_{\bot}^2(0)}{t^2}}
\label{eq2}
\end{equation}
where $R$ is the radius of the grains and $t$ is a characteristic length which measures the Josephson coupling between them. In the ultra-weak coupling limit, $t$ is infinite and Eq.\ref{eq2} reduces to the standard upper critical field of an individual spherical object\cite{Degennes64}. Here, the cross-section of the filaments is not a circle but an ellipse. In a first approximation, the previous formula applies if the radius $R$ is replaced by the half-axis of the ellipse perpendicular to the field, $R_{\bot}$.

In order to determine the precise texture, the coherence lengths have to be determined first. The zero field linear extrapolation of the upper critical field for H//\textit{c*}, H$_{c2,c}\approx 0.2T$  is similar to the value obtained for homogeneous superconductivity in \pfsix \cite{Lee97} or \clofour \cite{Murata87} and leads to the determination of $\xi_{\bot,c}=\sqrt{\xi_a\xi_b}\approx 41nm$. However, our zero field extrapolation of the upper critical field for H//\textit{b'} which is obtained by extrapolating linearly the upper critical field line from its value at low fields, H$_{c2,b}\approx 0.9 T$ is much smaller than measurements reported in \clofour or \pfsix performed with a perfect alignment of the magnetic field \cite{Murata87, Oh04, Joo06, Lee97}. This is due to our experimental set-up which do not allow a perfect alignment of our sample with respect to the \textit{b'}-axis under pressure. We can deduce a rough estimate of the misalignment of our sample to be about $10^{\circ}$ assuming that the upper critical field is identical for \clofour and \reofour for H//\textit{b'}. This misalignment is not crucial and in the following and we will correct the measurements from this angle. For clarity, we will denote \textit{b''} our direction of measurement. From H$_{c2,b}= 4 T$ \cite{Joo06}, we get $\xi_{\bot,b'}=\sqrt{\xi_a\xi_c}\approx 9nm$. Using $\xi_a=100nm$ as in \clofour\cite{Murata87}, we obtain $\xi_b\approx 4.1nm$ and $\xi_c\approx 0.9nm$ which is smaller than the lattice parameter \textit{c}. In this sense, the estimated value of $\xi_c$ is compatible with the observation of Josephson vortices in \clofour for H perfectly aligned along the \textit{b'}-axis\cite{Mansky95}. The misalignment of the field with respect to the \textit{b'}-axis avoids then to have a mixing of two different physical phenomena. Indeed, the possibility of Josephson vortices between the individual layers, separated by the lattice parameter \textit{c}, can be ruled out to analyze our data.

We now turn to the determination of the texture at P=10kbar. For simplicity, in absence of any X-ray data, we will assume that filaments elongated along the a-axis form a rectangular lattice in the \textit{b'-c} plane following the Deutscher \textit{et al.} model \cite{Deutscher80}. This assumption is justified by both the linear variation at low fields and the square root variation of the upper critical field and its observations along both directions. As a result, the distance between the filaments and the cross-section of each filament are nearly constant over the crystal. We will denote B and C, the distances between the filaments along \textit{b'} and \textit{c*} respectively. These quantities can be extracted from the H$_{c2}$ lines and using Eq.\ref{eq1} with $d_{\bot}$=B or C. To define T*, we use the intersection of the temperature axis with the parabolic fit of the $H_{c2}$ line at low temperatures which gives a more precise determination than any cross-over at a finite field. At P=10kbar, we get T*=0.75K and T*=0.85K along \textit{c*} and \textit{b''}-axis respectively which leads to B$\approx$80nm and C$\approx$50nm using only one significant digit.

\begin{figure}[htb]
	\centering
		\includegraphics[width=0.4\textwidth]{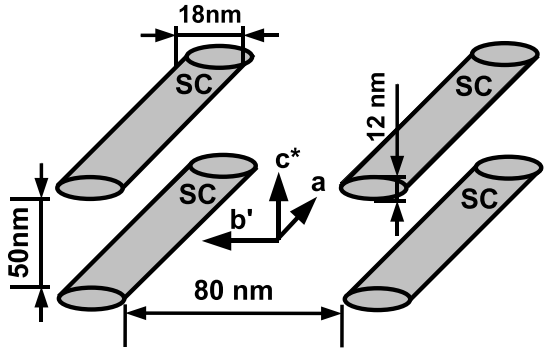}
	\caption{Texture of \reofour at P=10kbar deduced from the upper critical field data and assuming a rectangular lattice. SC denotes the superconducting filaments.}
	\label{texture}
\end{figure}

Finally, we can estimate the cross-section of the filaments at P=10kbar. In a first step, the coupling between the filaments can be assumed to be very small. From the fits of the low-temperature $H_{c2}$ lines, we can estimate $H_{c2,c}(0)\approx 2.5T $ and $H_{c2,b''}(0)\approx 2.9T$ so $H_{c2,b'}(0)\approx 17T$ from which we extract $R_{\bot,c}\approx 9nm $ which is about 2 times $\xi_b$ and $R_{\bot,b'}\approx 6nm$ which is about 6 times $\xi_c$ with a large uncertainty in this case. Indeed, we believe that for symmetry reasons, the ratio $R_{\bot,c}/\xi_b$ and $R_{\bot,b}/\xi_c$ should be identical. The discrepancy is attributed to the uncertainties in the determinations of both T* and $H_{c2}$. The texture proposed at P=10kbar is depicted on Fig.\ref{texture} assuming a rectangular lattice. As $R_{\bot,c}/ B \approx 1/9$ and $R_{\bot,b}/ C \approx 1/8$, the hypothesis of a low coupling between the filaments is justified a posteriori. The metallic fraction obtained is about 0.8\% in excellent agreement with the resistivity data \cite{Colin06} considering the crude assumptions made here. We notice that the cross-sections of the filaments are of the same order of magnitude as the coherence lengths justifying the term of 'mesoscopic phase separation'. In order to confirm our image, the evolution of the phenomena with increasing pressure has to be clarified. The basic influence of pressure is to increase the amount of metallic part which means larger filaments with a smaller distance between them. Smaller B and C parameters manifest by a decreasing temperature T*. This is effectively observed experimentally. As shown in Fig.\ref{Hc2}, from 10 to 11 kbar, T* varies from 0.75K to 0.2K leading to a decrease of the distance between the filaments from 80 to 60nm. This shift of T* is accompanied by a decrease of $H_{c2}$ at zero temperature which implies larger filaments and/or an increased coupling due to a lower distance between the filaments. The hypothesis of low coupling seems no more correct as we would obtain $R_{\bot,c}\approx 25nm$ which is now 40\% of the inter-filament distance along \textit{b'}. 

In summary, the superconducting data along two field directions has allowed the determination of the texture of a compound,\reofour, in its phase separation regime. The presence of both a linear regime and a parabolic law in the upper critical field curve have demonstrated the regular ordering of the individual filaments. This is a clear signature of a self-organization of the charge and that the physics studied here is not dominated by defects but only by long range interactions between the domains with a precise orientation of the anions. The data also demonstrate that phase separation occurs at the mesoscopic scale.

The authors acknowledge D.J\'erome for fruitful discussions. We are also grateful to P.Auban-Senzier for her assistance for the pressure equipment. This work is supported by the European Community under Grant COMEPHS No. NMPT4-CT-2005-517039. C.C. and B.S also acknowledge this grant for financial support.


\begin{thebibliography}{25}
\expandafter\ifx\csname natexlab\endcsname\relax\def\natexlab#1{#1}\fi
\expandafter\ifx\csname bibnamefont\endcsname\relax
  \def\bibnamefont#1{#1}\fi
\expandafter\ifx\csname bibfnamefont\endcsname\relax
  \def\bibfnamefont#1{#1}\fi
\expandafter\ifx\csname citenamefont\endcsname\relax
  \def\citenamefont#1{#1}\fi
\expandafter\ifx\csname url\endcsname\relax
  \def\url#1{\texttt{#1}}\fi
\expandafter\ifx\csname urlprefix\endcsname\relax\def\urlprefix{URL }\fi
\providecommand{\bibinfo}[2]{#2}
\providecommand{\eprint}[2][]{\url{#2}}

\bibitem[{\citenamefont{Joo et~al.}(2004)\citenamefont{Joo, Auban-Senzier,
  Pasquier, Monod., J\'erome, and Bechgaard}}]{Joo04}
\bibinfo{author}{\bibfnamefont{N.}~\bibnamefont{Joo}},
  \bibinfo{author}{\bibfnamefont{P.}~\bibnamefont{Auban-Senzier}},
  \bibinfo{author}{\bibfnamefont{C.~R.} \bibnamefont{Pasquier}},
  \bibinfo{author}{\bibfnamefont{P.}~\bibnamefont{Monod.}},
  \bibinfo{author}{\bibfnamefont{D.}~\bibnamefont{J\'erome}}, \bibnamefont{and}
  \bibinfo{author}{\bibfnamefont{K.}~\bibnamefont{Bechgaard}},
  \bibinfo{journal}{Eur. Phys. Journ. B} \textbf{\bibinfo{volume}{40}},
  \bibinfo{pages}{43} (\bibinfo{year}{2004}).

\bibitem[{\citenamefont{Joo et~al.}(2005)\citenamefont{Joo, Auban-Senzier,
  Pasquier, J\'erome, and Bechgaard}}]{Joo05}
\bibinfo{author}{\bibfnamefont{N.}~\bibnamefont{Joo}},
  \bibinfo{author}{\bibfnamefont{P.}~\bibnamefont{Auban-Senzier}},
  \bibinfo{author}{\bibfnamefont{C.~R.} \bibnamefont{Pasquier}},
  \bibinfo{author}{\bibfnamefont{D.}~\bibnamefont{J\'erome}}, \bibnamefont{and}
  \bibinfo{author}{\bibfnamefont{K.}~\bibnamefont{Bechgaard}},
  \bibinfo{journal}{Eur. Phys. Lett.} \textbf{\bibinfo{volume}{72}},
  \bibinfo{pages}{645} (\bibinfo{year}{2005}).

\bibitem[{\citenamefont{Schwenk et~al.}(1984)\citenamefont{Schwenk, Andres, and
  Wudl}}]{Schwenk84}
\bibinfo{author}{\bibfnamefont{H.}~\bibnamefont{Schwenk}},
  \bibinfo{author}{\bibfnamefont{K.}~\bibnamefont{Andres}}, \bibnamefont{and}
  \bibinfo{author}{\bibfnamefont{F.}~\bibnamefont{Wudl}},
  \bibinfo{journal}{Phys. Rev. B} \textbf{\bibinfo{volume}{29}},
  \bibinfo{pages}{500} (\bibinfo{year}{1984}).

\bibitem[{\citenamefont{Pouget et~al.}(1990)\citenamefont{Pouget, Kagoshima,
  Tamegai, Nogami, Kubo, Nakajima, and Bechgaard}}]{Pouget90}
\bibinfo{author}{\bibfnamefont{J.-P.} \bibnamefont{Pouget}},
  \bibinfo{author}{\bibfnamefont{S.}~\bibnamefont{Kagoshima}},
  \bibinfo{author}{\bibfnamefont{T.}~\bibnamefont{Tamegai}},
  \bibinfo{author}{\bibfnamefont{Y.}~\bibnamefont{Nogami}},
  \bibinfo{author}{\bibfnamefont{K.}~\bibnamefont{Kubo}},
  \bibinfo{author}{\bibfnamefont{T.}~\bibnamefont{Nakajima}}, \bibnamefont{and}
  \bibinfo{author}{\bibfnamefont{K.}~\bibnamefont{Bechgaard}},
  \bibinfo{journal}{Journ. Phys. Soc. Jpn.} \textbf{\bibinfo{volume}{59}},
  \bibinfo{pages}{2036} (\bibinfo{year}{1990}).

\bibitem[{\citenamefont{Lee et~al.}(2002)\citenamefont{Lee, Naughton, and
  Chaikin}}]{Lee02}
\bibinfo{author}{\bibfnamefont{I.~J.} \bibnamefont{Lee}},
     \bibinfo{author}{\bibfnamefont{P.~M.}
  \bibnamefont{Chaikin}}, \bibnamefont{and} \bibinfo{author}{\bibfnamefont{M.~J.} \bibnamefont{Naughton}},
   \bibinfo{journal}{Phys. Rev. Lett.}
  \textbf{\bibinfo{volume}{88}}, \bibinfo{pages}{207002}
  (\bibinfo{year}{2002}).

\bibitem[{\citenamefont{Vuleti\'{c} et~al.}(2002)\citenamefont{Vuleti\'{c},
  Auban-Senzier, Pasquier, Tomi\'{c}, J\'{e}rome, H\'{e}ritier, and
  Bechgaard}}]{Vuletic02}
\bibinfo{author}{\bibfnamefont{T.}~\bibnamefont{Vuleti\'{c}}},
  \bibinfo{author}{\bibfnamefont{P.}~\bibnamefont{Auban-Senzier}},
  \bibinfo{author}{\bibfnamefont{C.}~\bibnamefont{Pasquier}},
  \bibinfo{author}{\bibfnamefont{S.}~\bibnamefont{Tomi\'{c}}},
  \bibinfo{author}{\bibfnamefont{D.}~\bibnamefont{J\'{e}rome}},
  \bibinfo{author}{\bibfnamefont{M.}~\bibnamefont{H\'{e}ritier}},
  \bibnamefont{and}
  \bibinfo{author}{\bibfnamefont{K.}~\bibnamefont{Bechgaard}},
  \bibinfo{journal}{Eur. Phys. Journ. B} \textbf{\bibinfo{volume}{25}},
  \bibinfo{pages}{319} (\bibinfo{year}{2002}).

\bibitem[{\citenamefont{Colin et~al.}(2006)\citenamefont{Colin, Auban-Senzier,
  Pasquier, and Bechgaard}}]{Colin06}
\bibinfo{author}{\bibfnamefont{C.}~\bibnamefont{Colin}},
  \bibinfo{author}{\bibfnamefont{P.}~\bibnamefont{Auban-Senzier}},
  \bibinfo{author}{\bibfnamefont{C.~R.} \bibnamefont{Pasquier}},
  \bibnamefont{and}
  \bibinfo{author}{\bibfnamefont{K.}~\bibnamefont{Bechgaard}},
  \bibinfo{journal}{Eur. Phys. Lett.} \textbf{\bibinfo{volume}{75}},
  \bibinfo{pages}{301} (\bibinfo{year}{2006}).

\bibitem[{\citenamefont{Moret et~al.}(1982)\citenamefont{Moret, Pouget,
  Com\`es, and Bechgaard}}]{Moret82}
\bibinfo{author}{\bibfnamefont{R.}~\bibnamefont{Moret}},
  \bibinfo{author}{\bibfnamefont{J.-P.} \bibnamefont{Pouget}},
  \bibinfo{author}{\bibfnamefont{R.}~\bibnamefont{Com\`es}}, \bibnamefont{and}
  \bibinfo{author}{\bibfnamefont{K.}~\bibnamefont{Bechgaard}},
  \bibinfo{journal}{Phys. Rev. Lett.} \textbf{\bibinfo{volume}{49}},
  \bibinfo{pages}{1008} (\bibinfo{year}{1982}).

\bibitem[{\citenamefont{Turkevich and Klemm}(1979)}]{Turkevich79}
\bibinfo{author}{\bibfnamefont{L.A.}~\bibnamefont{Turkevich}} \bibnamefont{and}
  \bibinfo{author}{\bibfnamefont{R.A.}~\bibnamefont{Klemm}},
  \bibinfo{journal}{Phys. Rev. B} \textbf{\bibinfo{volume}{19}},
  \bibinfo{pages}{2520} (\bibinfo{year}{1979}).

\bibitem[{\citenamefont{Klemm et~al.}(1975)\citenamefont{Klemm, Luther, and
  Beasley}}]{Klemm75}
\bibinfo{author}{\bibfnamefont{R.}~\bibnamefont{Klemm}},
  \bibinfo{author}{\bibfnamefont{A.}~\bibnamefont{Luther}}, \bibnamefont{and}
  \bibinfo{author}{\bibfnamefont{M.}~\bibnamefont{Beasley}},
  \bibinfo{journal}{Phys. Rev. B.} \textbf{\bibinfo{volume}{12}},
  \bibinfo{pages}{877} (\bibinfo{year}{1975}).

\bibitem[{\citenamefont{Krasnov et~al.}(1996)\citenamefont{Krasnov, Kovalev,
  Oboznov, and Pedersen}}]{Krasnov96}
\bibinfo{author}{\bibfnamefont{V.~M.} \bibnamefont{Krasnov}},
  \bibinfo{author}{\bibfnamefont{A.~E.} \bibnamefont{Kovalev}},
  \bibinfo{author}{\bibfnamefont{V.~A.} \bibnamefont{Oboznov}},
  \bibnamefont{and} \bibinfo{author}{\bibfnamefont{N.~F.}
  \bibnamefont{Pedersen}}, \bibinfo{journal}{Phys. Rev. B}
  \textbf{\bibinfo{volume}{54}}, \bibinfo{pages}{15448} (\bibinfo{year}{1996}).

\bibitem[{\citenamefont{Deutscher et~al.}(1980)\citenamefont{Deutscher,
  Entin-Wohlman, and Shapira}}]{Deutscher80}
\bibinfo{author}{\bibfnamefont{G.}~\bibnamefont{Deutscher}},
  \bibinfo{author}{\bibfnamefont{O.}~\bibnamefont{Entin-Wohlman}},
  \bibnamefont{and} \bibinfo{author}{\bibfnamefont{Y.}~\bibnamefont{Shapira}},
  \bibinfo{journal}{Phys. Rev. B} \textbf{\bibinfo{volume}{22}},
  \bibinfo{pages}{4264} (\bibinfo{year}{1980}).

\bibitem[{\citenamefont{Parkin et~al.}(1981)\citenamefont{Parkin, J\'erome, and
  Bechgaard}}]{Parkin81}
\bibinfo{author}{\bibfnamefont{S.~S.~P.} \bibnamefont{Parkin}},
  \bibinfo{author}{\bibfnamefont{D.}~\bibnamefont{J\'erome}}, \bibnamefont{and}
  \bibinfo{author}{\bibfnamefont{K.}~\bibnamefont{Bechgaard}},
  \bibinfo{journal}{Mol. Cryst. Liq. Cryst.} \textbf{\bibinfo{volume}{79}},
  \bibinfo{pages}{213} (\bibinfo{year}{1981}).

\bibitem[{\citenamefont{Greene et~al.}(1986)\citenamefont{Greene, Parkin, and
  Schwenk}}]{Greene86}
\bibinfo{author}{\bibfnamefont{R.}~\bibnamefont{Greene}},
  \bibinfo{author}{\bibfnamefont{S.~S.~P.} \bibnamefont{Parkin}},
  \bibnamefont{and} \bibinfo{author}{\bibfnamefont{H.}~\bibnamefont{Schwenk}},
  \bibinfo{journal}{Physica B} \textbf{\bibinfo{volume}{143}},
  \bibinfo{pages}{388} (\bibinfo{year}{1986}).

\bibitem[{\citenamefont{Feinberg and Villard}(1990)}]{Feinberg90}
\bibinfo{author}{\bibfnamefont{D.}~\bibnamefont{Feinberg}} \bibnamefont{and}
  \bibinfo{author}{\bibfnamefont{C.}~\bibnamefont{Villard}},
  \bibinfo{journal}{Phys. Rev. Lett.} \textbf{\bibinfo{volume}{65}},
  \bibinfo{pages}{919} (\bibinfo{year}{1990}).

\bibitem[{\citenamefont{Lawrence and Doniach}(1971)}]{LD}
\bibinfo{author}{\bibfnamefont{W.}~\bibnamefont{Lawrence}} \bibnamefont{and}
  \bibinfo{author}{\bibfnamefont{S.}~\bibnamefont{Doniach}}, in
  \emph{\bibinfo{booktitle}{Proceedings of the 12th International Conference on
  Low Temperature Physics}}, edited by
  \bibinfo{editor}{\bibfnamefont{E.}~\bibnamefont{Kanda}}
  (\bibinfo{year}{1971}), p. \bibinfo{pages}{361}.

\bibitem[{\citenamefont{Fulde and Ferrell}(1964)}]{Fulde64}
\bibinfo{author}{\bibfnamefont{P.}~\bibnamefont{Fulde}} \bibnamefont{and}
  \bibinfo{author}{\bibfnamefont{R.~A.} \bibnamefont{Ferrell}},
  \bibinfo{journal}{Phys.Rev.} \textbf{\bibinfo{volume}{135}},
  \bibinfo{pages}{A550} (\bibinfo{year}{1964}).

\bibitem[{\citenamefont{Larkin and Ovchinnikov}(1964)}]{Larkin64}
\bibinfo{author}{\bibfnamefont{A.~I.} \bibnamefont{Larkin}} \bibnamefont{and}
  \bibinfo{author}{\bibfnamefont{Y.~N.} \bibnamefont{Ovchinnikov}},
  \bibinfo{journal}{Zh.Eksp.Teor.Fiz.} \textbf{\bibinfo{volume}{47}},
  \bibinfo{pages}{113} (\bibinfo{year}{1964}), \bibinfo{note}{\ Sov. Phys. JETP
  \textbf{20},762 (1965)}.

\bibitem[{\citenamefont{N.Dupuis}(1995)}]{Dupuis95}
\bibinfo{author}{\bibnamefont{N.Dupuis}}, \bibinfo{journal}{Phys. Rev. B}
  \textbf{\bibinfo{volume}{51}}, \bibinfo{pages}{9074} (\bibinfo{year}{1995}).

\bibitem[{\citenamefont{de~Gennes and Tinkham}(1964)}]{Degennes64}
\bibinfo{author}{\bibfnamefont{P.}~\bibnamefont{de~Gennes}} \bibnamefont{and}
  \bibinfo{author}{\bibfnamefont{M.}~\bibnamefont{Tinkham}},
  \bibinfo{journal}{Physics (N.Y.)} \textbf{\bibinfo{volume}{1}},
  \bibinfo{pages}{107} (\bibinfo{year}{1964}).

\bibitem[{\citenamefont{Lee et~al.}(1997)\citenamefont{Lee, Naughton, Danner,
  and Chaikin}}]{Lee97}
\bibinfo{author}{\bibfnamefont{I.~J.} \bibnamefont{Lee}},
  \bibinfo{author}{\bibfnamefont{M.~J.} \bibnamefont{Naughton}},
  \bibinfo{author}{\bibfnamefont{G.~M.} \bibnamefont{Danner}},
  \bibnamefont{and} \bibinfo{author}{\bibfnamefont{P.~M.}
  \bibnamefont{Chaikin}}, \bibinfo{journal}{Phys. Rev. Lett.}
  \textbf{\bibinfo{volume}{78}}, \bibinfo{pages}{3555} (\bibinfo{year}{1997}).

\bibitem[{\citenamefont{Murata et~al.}(1987)\citenamefont{Murata, Tokumoto,
  Anzai, Kajimura, and Ishiguro}}]{Murata87}
\bibinfo{author}{\bibfnamefont{K.}~\bibnamefont{Murata}},
  \bibinfo{author}{\bibfnamefont{M.}~\bibnamefont{Tokumoto}},
  \bibinfo{author}{\bibfnamefont{H.}~\bibnamefont{Anzai}},
  \bibinfo{author}{\bibfnamefont{K.}~\bibnamefont{Kajimura}}, \bibnamefont{and}
  \bibinfo{author}{\bibfnamefont{T.}~\bibnamefont{Ishiguro}},
  \bibinfo{journal}{Jpn. Journ. of Appl. Physics}
  \textbf{\bibinfo{volume}{Suppl. 26-3}}, \bibinfo{pages}{1367}
  (\bibinfo{year}{1987}).

\bibitem[{\citenamefont{Oh and Naughton}(2004)}]{Oh04}
\bibinfo{author}{\bibfnamefont{J.~I.} \bibnamefont{Oh}} \bibnamefont{and}
  \bibinfo{author}{\bibfnamefont{M.~J.} \bibnamefont{Naughton}},
  \bibinfo{journal}{Phys. Rev. Lett.} \textbf{\bibinfo{volume}{92}},
  \bibinfo{pages}{67001} (\bibinfo{year}{2004}).

\bibitem[{\citenamefont{Joo et~al.}(2006)\citenamefont{Joo, Auban-Senzier,
  Pasquier, Yonezawa, Higashinnaka, Maeno, Haddad, Charfi-Kaddour, H\'eritier,
  Bechgaard et~al.}}]{Joo06}
\bibinfo{author}{\bibfnamefont{N.}~\bibnamefont{Joo}},
  \bibinfo{author}{\bibfnamefont{P.}~\bibnamefont{Auban-Senzier}},
  \bibinfo{author}{\bibfnamefont{C.~R.} \bibnamefont{Pasquier}},
  \bibinfo{author}{\bibfnamefont{S.}~\bibnamefont{Yonezawa}},
  \bibinfo{author}{\bibfnamefont{H.}~\bibnamefont{Higashinnaka}},
  \bibinfo{author}{\bibfnamefont{Y.}~\bibnamefont{Maeno}},
  \bibinfo{author}{\bibfnamefont{S.}~\bibnamefont{Haddad}},
  \bibinfo{author}{\bibfnamefont{S.}~\bibnamefont{Charfi-Kaddour}},
  \bibinfo{author}{\bibfnamefont{M.}~\bibnamefont{H\'eritier}},
  \bibinfo{author}{\bibfnamefont{K.}~\bibnamefont{Bechgaard}},
  \bibnamefont{et~al.}, \bibinfo{journal}{Eur. Phys. Journ. B}
  \textbf{\bibinfo{volume}{52}}, \bibinfo{pages}{337} (\bibinfo{year}{2006}).

\bibitem[{\citenamefont{Mansky et~al.}(1995)\citenamefont{Mansky, Danner, and
  Chaikin}}]{Mansky95}
\bibinfo{author}{\bibfnamefont{P.~A.} \bibnamefont{Mansky}},
  \bibinfo{author}{\bibfnamefont{G.}~\bibnamefont{Danner}}, \bibnamefont{and}
  \bibinfo{author}{\bibfnamefont{P.~M.} \bibnamefont{Chaikin}},
  \bibinfo{journal}{Phys. Rev. B} \textbf{\bibinfo{volume}{52}},
  \bibinfo{pages}{7554} (\bibinfo{year}{1995}).

\end{thebibliography}

\end{document}